\documentstyle[10pt,epsf,epsfig,dp_delphititle,oldlfont,amssymb]{dp_delphi}
\makeindex
\def\DpPaperGroup{PH--EP}
\def\DpPaperRef{2008-017}
\def\DpDate{10 November 2008}
\def\DpAuthors{DELPHI Collaboration}
\def\DpTitle{{
 Inclusive single-particle production in two-photon collisions 
at LEP II with the DELPHI detector
}}
\def\DpSubmit{(Accepted by Phys. Lett. B)}
\def\DpComment{}
\def\DpEMail{}
\def\jpsi{\hbox{$J{\kern-0.24em}/{\kern-0.14em}\psi$}}
\def\ev#1#2{\hbox{#1e{\kern-0.10em}V{\kern-0.30em}/{\kern-0.14em}$#2$}}

\def\bqt#1#2\eqt{\begin{equation}\label{#1}%
        {#2}\end{equation}\noindent}
\def\bln#1#2\eln{\begin{equation}\label{#1}%
            \eqalign{#2}\end{equation}\noindent}
\def\brlist{\begin{thebibliography}{99}}

\def\erlist{\end{thebibliography}}

\begin{document}
\makeatletter
\newcount\@tempcntc
\def\@citex[#1]#2{\if@filesw\immediate\write\@auxout{\string\citation{#2}}\fi
  \@tempcnta\z@\@tempcntb\m@ne\def\@citea{}\@cite{\@for\@citeb:=#2\do
    {\@ifundefined
       {b@\@citeb}{\@citeo\@tempcntb\m@ne\@citea\def\@citea{,}{\bf ?}\@warning
       {Citation `\@citeb' on page \thepage \space undefined}}%
    {\setbox\z@\hbox{\global\@tempcntc0\csname b@\@citeb\endcsname\relax}%
     \ifnum\@tempcntc=\z@ \@citeo\@tempcntb\m@ne
       \@citea\def\@citea{,}\hbox{\csname b@\@citeb\endcsname}%
     \else
      \advance\@tempcntb\@ne
      \ifnum\@tempcntb=\@tempcntc
      \else\advance\@tempcntb\m@ne\@citeo
      \@tempcnta\@tempcntc\@tempcntb\@tempcntc\fi\fi}}\@citeo}{#1}}
\def\@citeo{\ifnum\@tempcnta>\@tempcntb\else\@citea\def\@citea{,}%
  \ifnum\@tempcnta=\@tempcntb\the\@tempcnta\else
   {\advance\@tempcnta\@ne\ifnum\@tempcnta=\@tempcntb \else \def\@citea{--}\fi
    \advance\@tempcnta\m@ne\the\@tempcnta\@citea\the\@tempcntb}\fi\fi}
 
\makeatother

\begin{titlepage}
\pagenumbering{roman}

\CERNpreprint{\DpPaperGroup}{\DpPaperRef}   
\date{{\small\DpDate}}			    
\title{\DpTitle}			    
\address{\DpAuthors}			    

\begin{shortabs}			    
\noindent
A study of the inclusive charged hadron production in two-photon collisions
is described. The data were collected with the DELPHI detector at LEP II. 
Results on the inclusive single-particle $p_T$ distribution and 
the differential charged hadrons $d\sigma/dp_T$ cross-section are presented 
and compared to the predictions of perturbative NLO QCD calculations 
and to published results.
\end{shortabs}

\vfill

\begin{center}
\DpSubmit \ \\		
\DpComment \ \\
\DpEMail \ \\
\end{center}

\vfill
\clearpage

\headsep 10.0pt

\addtolength{\textheight}{10mm}
\addtolength{\footskip}{-5mm}
\begingroup
%
\newcommand{\DpName}[2]{\hbox{#1$^{\ref{#2}}$},\hfill}
\newcommand{\DpNameTwo}[3]{\hbox{#1$^{\ref{#2},\ref{#3}}$},\hfill}
\newcommand{\DpNameThree}[4]{\hbox{#1$^{\ref{#2},\ref{#3},\ref{#4}}$},\hfill}
\newskip\Bigfill \Bigfill = 0pt plus 1000fill
\newcommand{\DpNameLast}[2]{\hbox{#1$^{\ref{#2}}$}\hspace{\Bigfill}}

%
\footnotesize
\noindent
\DpName{J.Abdallah}{LPNHE}
\DpName{P.Abreu}{LIP}
\DpName{W.Adam}{VIENNA}
\DpName{P.Adzic}{DEMOKRITOS}
\DpName{T.Albrecht}{KARLSRUHE}
\DpName{R.Alemany-Fernandez}{CERN}
\DpName{T.Allmendinger}{KARLSRUHE}
\DpName{P.P.Allport}{LIVERPOOL}
\DpName{U.Amaldi}{MILANO2}
\DpName{N.Amapane}{TORINO}
% ...
\normalsize
\endgroup

\newpage
\titlefoot{Department of Physics and Astronomy, Iowa State
     University, Ames IA 50011-3160, USA
    \label{AMES}}
\titlefoot{Physics Department, Universiteit Antwerpen,
     Universiteitsplein 1, B-2610 Antwerpen, Belgium
    \label{ANTWERP}}
\titlefoot{IIHE, ULB-VUB,
     Pleinlaan 2, B-1050 Brussels, Belgium
    \label{BRUSSELS}}
\titlefoot{Physics Laboratory, University of Athens, Solonos Str.
     104, GR-10680 Athens, Greece
    \label{ATHENS}}
\titlefoot{Department of Physics, University of Bergen,
     All\'egaten 55, NO-5007 Bergen, Norway
    \label{BERGEN}}
\titlefoot{Dipartimento di Fisica, Universit\`a di Bologna and INFN,
     Viale C. Berti Pichat 6/2, IT-40127 Bologna, Italy
    \label{BOLOGNA}}
\titlefoot{Centro Brasileiro de Pesquisas F\'{\i}sicas, rua Xavier Sigaud 150,
     BR-22290 Rio de Janeiro, Brazil
    \label{BRASIL-CBPF}}
\titlefoot{Inst. de F\'{\i}sica, Univ. Estadual do Rio de Janeiro,
     rua S\~{a}o Francisco Xavier 524, Rio de Janeiro, Brazil
    \label{BRASIL-IFUERJ}}
\titlefoot{Coll\`ege de France, Lab. de Physique Corpusculaire, IN2P3-CNRS,
     FR-75231 Paris Cedex 05, France
    \label{CDF}}
\titlefoot{CERN, CH-1211 Geneva 23, Switzerland
    \label{CERN}}
\titlefoot{Institut de Recherches Subatomiques, IN2P3 - CNRS/ULP - BP20,
     FR-67037 Strasbourg Cedex, France
    \label{CRN}}
\titlefoot{Now at DESY-Zeuthen, Platanenallee 6, D-15735 Zeuthen, Germany
    \label{DESY}}
\titlefoot{Institute of Nuclear Physics, N.C.S.R. Demokritos,
     P.O. Box 60228, GR-15310 Athens, Greece
    \label{DEMOKRITOS}}
\titlefoot{FZU, Inst. of Phys. of the C.A.S. High Energy Physics Division,
     Na Slovance 2, CZ-182 21, Praha 8, Czech Republic
    \label{FZU}}
\titlefoot{Dipartimento di Fisica, Universit\`a di Genova and INFN,
     Via Dodecaneso 33, IT-16146 Genova, Italy
    \label{GENOVA}}
\titlefoot{Institut des Sciences Nucl\'eaires, IN2P3-CNRS, Universit\'e
     de Grenoble 1, FR-38026 Grenoble Cedex, France
    \label{GRENOBLE}}
\titlefoot{Helsinki Institute of Physics and Department of Physical Sciences,
     P.O. Box 64, FIN-00014 University of Helsinki, 
     \indent~~Finland
    \label{HELSINKI}}
\titlefoot{Joint Institute for Nuclear Research, Dubna, Head Post
     Office, P.O. Box 79, RU-101 000 Moscow, Russian Federation
    \label{JINR}}
\titlefoot{Institut f\"ur Experimentelle Kernphysik,
     Universit\"at Karlsruhe, Postfach 6980, DE-76128 Karlsruhe,
     Germany
    \label{KARLSRUHE}}
\titlefoot{Institute of Nuclear Physics PAN,Ul. Radzikowskiego 152,
     PL-31142 Krakow, Poland
    \label{KRAKOW1}}
\titlefoot{Faculty of Physics and Nuclear Techniques, University of Mining
     and Metallurgy, PL-30055 Krakow, Poland
    \label{KRAKOW2}}
\titlefoot{LAL, Univ Paris-Sud, CNRS/IN2P3, Orsay, France
    \label{LAL}}
\titlefoot{School of Physics and Chemistry, University of Lancaster,
     Lancaster LA1 4YB, UK
    \label{LANCASTER}}
\titlefoot{LIP, IST, FCUL - Av. Elias Garcia, 14-$1^{o}$,
     PT-1000 Lisboa Codex, Portugal
    \label{LIP}}
\titlefoot{Department of Physics, University of Liverpool, P.O.
     Box 147, Liverpool L69 3BX, UK
    \label{LIVERPOOL}}
\titlefoot{Dept. of Physics and Astronomy, Kelvin Building,
     University of Glasgow, Glasgow G12 8QQ, UK
    \label{GLASGOW}}
\titlefoot{LPNHE, IN2P3-CNRS, Univ.~Paris VI et VII, Tour 33 (RdC),
     4 place Jussieu, FR-75252 Paris Cedex 05, France
    \label{LPNHE}}
\titlefoot{Department of Physics, University of Lund,
     S\"olvegatan 14, SE-223 63 Lund, Sweden
    \label{LUND}}
\titlefoot{Universit\'e Claude Bernard de Lyon, IPNL, IN2P3-CNRS,
     FR-69622 Villeurbanne Cedex, France
    \label{LYON}}
\titlefoot{Dipartimento di Fisica, Universit\`a di Milano and INFN-MILANO,
     Via Celoria 16, IT-20133 Milan, Italy
    \label{MILANO}}
\titlefoot{Dipartimento di Fisica, Univ. di Milano-Bicocca and
     INFN-MILANO, Piazza della Scienza 3, IT-20126 Milan, Italy
    \label{MILANO2}}
\titlefoot{IPNP of MFF, Charles Univ., Areal MFF,
     V Holesovickach 2, CZ-180 00, Praha 8, Czech Republic
    \label{NC}}
\titlefoot{NIKHEF, Postbus 41882, NL-1009 DB
     Amsterdam, The Netherlands
    \label{NIKHEF}}
\titlefoot{National Technical University, Physics Department,
     Zografou Campus, GR-15773 Athens, Greece
    \label{NTU-ATHENS}}
\titlefoot{Physics Department, University of Oslo, Blindern,
     NO-0316 Oslo, Norway
    \label{OSLO}}
\titlefoot{Dpto. Fisica, Univ. Oviedo, Avda. Calvo Sotelo
     s/n, ES-33007 Oviedo, Spain
    \label{OVIEDO}}
\titlefoot{Department of Physics, University of Oxford,
     Keble Road, Oxford OX1 3RH, UK
    \label{OXFORD}}
\titlefoot{Dipartimento di Fisica, Universit\`a di Padova and
     INFN, Via Marzolo 8, IT-35131 Padua, Italy
    \label{PADOVA}}
\titlefoot{Rutherford Appleton Laboratory, Chilton, Didcot
     OX11 OQX, UK
    \label{RAL}}
\titlefoot{Dipartimento di Fisica, Universit\`a di Roma II and
     INFN, Tor Vergata, IT-00173 Rome, Italy
    \label{ROMA2}}
\titlefoot{Dipartimento di Fisica, Universit\`a di Roma III and
     INFN, Via della Vasca Navale 84, IT-00146 Rome, Italy
    \label{ROMA3}}
\titlefoot{DAPNIA/Service de Physique des Particules,
     CEA-Saclay, FR-91191 Gif-sur-Yvette Cedex, France
    \label{SACLAY}}
\titlefoot{Instituto de Fisica de Cantabria (CSIC-UC), Avda.
     los Castros s/n, ES-39006 Santander, Spain
    \label{SANTANDER}}
\titlefoot{Inst. for High Energy Physics, Serpukov
     P.O. Box 35, Protvino, (Moscow Region), Russian Federation
    \label{SERPUKHOV}}
\titlefoot{J. Stefan Institute, Jamova 39, SI-1000 Ljubljana, Slovenia
    \label{SLOVENIJA1}}
\titlefoot{Laboratory for Astroparticle Physics,
     University of Nova Gorica, Kostanjeviska 16a, SI-5000 Nova Gorica, Slovenia
    \label{SLOVENIJA2}}
\titlefoot{Department of Physics, University of Ljubljana,
     SI-1000 Ljubljana, Slovenia
    \label{SLOVENIJA3}}
\titlefoot{Fysikum, Stockholm University,
     Box 6730, SE-113 85 Stockholm, Sweden
    \label{STOCKHOLM}}
\titlefoot{Dipartimento di Fisica Sperimentale, Universit\`a di
     Torino and INFN, Via P. Giuria 1, IT-10125 Turin, Italy
    \label{TORINO}}
\titlefoot{INFN,Sezione di Torino and Dipartimento di Fisica Teorica,
     Universit\`a di Torino, Via Giuria 1,
     IT-10125 Turin, Italy
    \label{TORINOTH}}
\titlefoot{Dipartimento di Fisica, Universit\`a di Trieste and
     INFN, Via A. Valerio 2, IT-34127 Trieste, Italy
    \label{TRIESTE}}
\titlefoot{Istituto di Fisica, Universit\`a di Udine and INFN,
     IT-33100 Udine, Italy
    \label{UDINE}}
\titlefoot{Univ. Federal do Rio de Janeiro, C.P. 68528
     Cidade Univ., Ilha do Fund\~ao
     BR-21945-970 Rio de Janeiro, Brazil
    \label{UFRJ}}
\titlefoot{Department of Radiation Sciences, University of
     Uppsala, P.O. Box 535, SE-751 21 Uppsala, Sweden
    \label{UPPSALA}}
\titlefoot{IFIC, Valencia-CSIC, and D.F.A.M.N., U. de Valencia,
     Avda. Dr. Moliner 50, ES-46100 Burjassot (Valencia), Spain
    \label{VALENCIA}}
\titlefoot{Institut f\"ur Hochenergiephysik, \"Osterr. Akad.
     d. Wissensch., Nikolsdorfergasse 18, AT-1050 Vienna, Austria
    \label{VIENNA}}
\titlefoot{Inst. Nuclear Studies and University of Warsaw, Ul.
     Hoza 69, PL-00681 Warsaw, Poland
    \label{WARSZAWA}}
\titlefoot{Now at University of Warwick, Coventry CV4 7AL, UK
    \label{WARWICK}}
\titlefoot{Fachbereich Physik, University of Wuppertal, Postfach
     100 127, DE-42097 Wuppertal, Germany \\
\noindent
{$^\dagger$~deceased}
    \label{WUPPERTAL}}
\addtolength{\textheight}{-10mm}
\addtolength{\footskip}{5mm}
\clearpage

\headsep 30.0pt
\end{titlepage}

%
\pagenumbering{arabic}				    
\setcounter{footnote}{0}			    %
\large
\section{Introduction}
The inclusive production of hadrons in $\gamma^*\gamma^*$ interactions can be 
used to study the structure of two-photon collisions ~\cite{GAMGAM}. These 
photons are radiated by beam electrons which scatter at very small angles and 
most of them are not detected. The untagged photons are quasi-real 
with a mass $Q^2\sim 0$.
 At LEP II these collisions are the main source of hadron production, providing 
a good opportunity for such an investigation and thus to check the predictions 
of leading and next-to-leading order (NLO) perturbative QCD computations.

The L3 and OPAL collaborations have published results of their analyses of 
the inclusive production of charged hadrons in two-photon collisions 
~\cite{L3,OPAL}. While L3 observes a pion production cross-section largely 
exceeding the NLO QCD predictions at high transverse momenta  
(5 GeV{\it/c} $< p_T <$~17 GeV{\it/c}), OPAL finds a good agreement 
with them, in the $p_T <$~10 GeV{\it/c} range of its analysis.

In this paper we present the DELPHI study of the inclusive production of 
charged hadrons in collisions of quasi-real photons. Section 2 describes the 
selection criteria for the event sample collected for this study. The 
inclusive single-particle transverse momentum spectrum and the measurement 
of the differential charged hadrons cross-section are presented in Section 3. 
They are compared to theoretical QCD predictions and published results in 
Section 4.

\section{Experimental procedure}
The analysis presented here is based on the data taken with the DELPHI 
detector~\cite{DELPHI1,DELPHI2} in 1996-2000, covering  
a range of centre-of-mass energies from 161 GeV to 209 GeV, with a  
luminosity-weighted average centre-of-mass energy: 195.5 GeV.
The selected data set corresponds to the period when 
 the Time Projection Chamber (TPC), the main tracking device of DELPHI,
was fully operational thus ensuring good particle reconstruction.
The corresponding integrated luminosity used in this analysis is
617 ${\mathrm p}{\mathrm b}^{-1}$.

The charged particles were measured in the tracking system of DELPHI, which
consists of the microVertex Detector (VD), the Inner Detector (ID),
the TPC, the Outer Detector (OD) in the barrel, and the
Forward Chambers FCA and FCB in the endcaps of DELPHI, all
embedded in a homogeneous 1.2 T magnetic field. The following selection 
criteria are applied to charged particles :
\begin{itemize}
\item transverse momentum $p_T \ >$ 150 MeV{\it/c};
\item impact parameter of a trajectory transverse to the beam axis
$\Delta_{xy} \ <$ 0.4 cm;
\item impact parameter of a trajectory along the beam axis
$\Delta_{z} \ <$ 2 cm;
\item polar angle of a track with respect to the e$^-$ 
beam $10^\circ<\theta<170^\circ$;
\item track length $l \ >$ 30 cm;
\item relative error of its momentum $\Delta p/p \ <$ 100\%.
\end{itemize}  

The measurement of neutral particles is made using the calorimeter information
provided by the electromagnetic calorimeters, the High Density Projection 
Chamber (HPC) in the barrel and Forward Electromagnetic Calorimeter (FEMC) in 
the forward (backward) regions and by the hadronic calorimeter (HAC). 
Events with photons tagged by the DELPHI luminometer (STIC),
 i.e. with high $Q^2$  values, have been rejected.
The calorimeter clusters, which are not associated to charged particle tracks, 
are combined to form the signals from the neutral particles 
($\gamma, \ \pi^0, \ {\mathrm K}^0_L,$ n). The following thresholds are set 
on the measured energy: 0.5~GeV for showers in the electromagnetic calorimeters 
and 2~GeV for showers in the hadronic calorimeter. Furthermore the polar angle
of neutral tracks was required to be in the range $10^\circ<\theta<170^\circ$.
 
To extract the hadronic events from the collisions of quasi-real photons the 
following cuts are applied:   
\begin{itemize}
\item  energy deposited in the DELPHI luminometer (STIC:
~$2.5^{\circ}<\theta_{STIC}<9^{\circ}$) \mbox{E$_{STIC}  \ <$ 30 GeV};
\item number of charged-particle tracks  
 $N_{ch} \ >$ 4;
\item visible invariant mass, calculated from the four-momentum vectors of 
the measured charged and neutral particles, assuming the pion mass for charged 
particles, \mbox{5 GeV{\it/$c^2$}$< W_{vis} < $ 35 GeV{\it/$c^2$}}.
\end{itemize}

The first condition eliminates the so-called single and double-tagged  
$\gamma^* \gamma^*$ events.
The condition on the charged track muliplicity as well as the lower limit on 
$W_{vis}$ reduce the background 
from $\gamma^* \gamma^* \rightarrow \tau^+ \tau^-$ events.
The upper limit on $W_{vis}$ cuts down the background from the
\mbox{e$^+$e$^- \rightarrow q \bar{q}~(\gamma)$},
e$^+$e$^- \rightarrow \tau^+ \tau^-$ and four-fermion processes. 
The comparison of the $W_{vis}$ distributions (Fig.~1) 
for the data and the Monte Carlo (MC) generated samples of events, described 
below, illustrates the effects of the $W_{vis}$ cuts.

About 910k events are selected after application of the above selection
criteria.

\section{Data Analysis and Results}
Monte Carlo samples of the various final states present in the data were
generated for comparison with these data. The simulation of the process 
\mbox{$\gamma^*\gamma^* \rightarrow \ hadrons$} was based on PYTHIA 6.143 
~\cite{PYTHIA} in which the description of the hadron production encompasses 
the processes described by the Quark Parton Model (QPM) (direct process), 
the Vector Dominance Model (VDM) and the hard scattering of the hadronic 
constituents of quasi-real photons (resolved photon process). The MC sample 
of events used is 2.7 times larger than the data. 
The main background coming from the inclusive 
\mbox{e$^+$e$^- \rightarrow q \bar{q}~(\gamma)$} channel has been estimated 
from a PYTHIA 6.125 sample. The simulations of the 
\mbox{${\mathrm e}^+{\mathrm e}^-\rightarrow$ four-fermion}, 
the  $\gamma^* \gamma^* \rightarrow \tau^+ \tau^-$   
and of the e$^+$e$^- \rightarrow \tau^+ \tau^-$ backgrounds were based on 
the EXCALIBUR ~\cite{EXCALIBUR}, BDKRC~\cite{BDKRS} and KORALZ 4.2 
~\cite{KORALZ} generators, respectively. 
The Monte Carlo generated events were then passed  through the standard
DELPHI detector simulation and reconstruction programs ~\cite{DELPHI2}.
The same cuts were applied on the reconstructed MC events as on the data. 

The $dN/dp_T$ distribution of the charged particles of the selected events is 
presented in \mbox{Fig. 2}, for tracks with pseudo-rapidity 
$|\eta|<$~1 ($\eta=-\ln~\rm{tan}(\theta/2)$) 
\footnote{The angular selection of tracks (Table 1 and Figs. 2-5) is 
expressed in terms of $|\eta|$ cuts for comparison with 
published results ~\cite{L3,OPAL}.}
, i.e. well measured tracks including TPC information. The expected Monte 
Carlo generated contributions, normalized to the data integrated luminosity are 
also shown. The data are well reproduced by the 
sum of the simulated samples of events for \mbox{$p_T >$~1.6 GeV{\it/c}} and 
the \mbox{e$^+$e$^- \rightarrow q \bar{q}~(\gamma)$} channel is the main 
contributor for $p_T >$~12 GeV{\it/c}. There is a lack of data at 
$p_T <$~1.6 GeV{\it/c}, becoming substantial at $p_T <$~1 GeV{\it/c}. This  
is caused by the trigger efficiency which was not accounted for in the Monte 
Carlo simulation and which is low for low $p_T$ tracks and low multiplicities
~\cite{TRIGGER}. For this reason, the following comparison with theoretical 
predictions is presented for $p_T >$~1.6 GeV{\it/c} only.

The differential $d\sigma/dp_T$ cross-section distribution of the inclusive
production of charged hadrons in the process 
\mbox{$\gamma^*\gamma^* \rightarrow \ hadrons$} has been obtained by 
subtracting the background contributions from the experimental $dN/dp_T$ data. 
The resulting distribution has been corrected, bin-by-bin, by a factor which 
is the inverse of the ratio of the numbers of reconstructed to generated tracks
of \mbox{$\gamma^*\gamma^* \rightarrow \ hadrons$} in Monte Carlo events. 
This ratio  is of the order of 50-60\% for 
\mbox{1.6 GeV{\it/c}~$< p_T < $~4 GeV{\it/c}} and drops to about 20\% for
$p_T >$~10 GeV{\it/c}, the upper bound on $W_{vis}$ being mainly responsible 
for the drop in efficiency on large $p_T$ tracks. The $d\sigma/dp_T$ distribution
is shown in Fig. 3 for different sets of selection 
criteria as described below. The PYTHIA prediction is also shown. It agrees 
very well with the data for $p_T >$~1.6 GeV{\it/c} ~up to large $p_T$ values.
 
To study the systematic uncertainty coming from the selection criteria, we
have varied them, in particular the $W_{vis}$ upper limit and the track polar 
angle $(\theta)$ cuts. A smaller upper bound of $W_{vis}$ has the advantage 
of minimizing the background contributions especially the 
\mbox{e$^+$e$^- \rightarrow q \bar{q}~(\gamma)$} one. Tracks at low polar 
angle are missing TPC measurements and are thus less well
measured. On the other hand most contributing processes correspond to the 
emission of tracks peaked in the forward (backward) regions, in particular
the \mbox{e$^+$e$^- \rightarrow q \bar{q}~(\gamma)$} and even more the
\mbox{$\gamma^*\gamma^* \rightarrow \ hadrons$}  channels. 
Hence a tight $(\theta)$ cut can reduce significantly the number of selected
charged-particle tracks ($N_{ch}$) of a given event and consequently its
computed visible energy $W_{vis}$.
\mbox{Fig. 3} shows the $d\sigma/dp_T$ distributions, calculated
using tracks with $|\eta|<$1.5, 
for four sets of 
selection criteria varying the polar angle selection imposed on 
charged tracks and the cut on the visible invariant mass $W_{vis}$:

\begin{enumerate}
\item   $10^{\circ} \ < \ \theta \ < \ 170^{\circ} ~~~(|\eta|<2.4)$, 
\mbox{~5 GeV{\it/$c^2$}$< W_{vis} < $20 GeV{\it/$c^2$}};
\item   $25^{\circ} \ < \ \theta \ < \ 155^{\circ} ~~~(|\eta|<1.5)$, 
\mbox{~5 GeV{\it/$c^2$}$< W_{vis} < $20 GeV{\it/$c^2$}};
\item   $10^{\circ} \ < \ \theta \ < \ 170^{\circ} ~~~(|\eta|<2.4)$, 
\mbox{~5 GeV{\it/$c^2$}$< W_{vis} < $35 GeV{\it/$c^2$}};
\item   $25^{\circ} \ < \ \theta \ < \ 155^{\circ} ~~~(|\eta|<1.5)$, 
\mbox{~5 GeV{\it/$c^2$}$< W_{vis} < $35 GeV{\it/$c^2$}}.

\end{enumerate}

The spread of the measurements is relatively small for 
$p_T <$ 7-8 GeV{\it/c} but increases for high $p_T$ values where the 
\mbox{e$^+$e$^- \rightarrow q \bar{q}~(\gamma)$} dominates. The corresponding
systematic uncertainty has been estimated as half of the spread of the four
sets of measurements.

The other source of uncertainty comes from the Monte Carlo modelling. It has 
been estimated by comparing the PYTHIA and TWOGAM~\cite{TWOGAM} predictions 
for the \mbox{$\gamma^*\gamma^* \rightarrow \ hadrons$} processes and PYTHIA 
and HERWIG~\cite{HERWIG} predictions for the 
\mbox{e$^+$e$^- \rightarrow q \bar{q}~(\gamma)$} process. It was found 
that the relative difference on the efficiencies calculated from the various
generators depends on $p_T$ but never exceeds 10\%. 
The corresponding uncertainty has been defined as half of the difference 
between two generator contributions. All systematic uncertainties have been 
added quadratically in Table 1. 
   
Table 1 gives the values of $d\sigma/dp_T$ as a function of $p_T$, for the 
selection criteria described in section 2, the pseudo-rapidity ranges 
$|\eta|<$~1 and  $|\eta|<$~1.5 and for $p_T>$~1.6 GeV{\it/c} where the 
event trigger efficiency is close to 100\%. The first 
error is statistical and the second one is the overall systematic uncertainty.
Fig.~4 shows the comparison of the $d\sigma/dp_T$ distribution for 
$|\eta|<~1.5$ with the NLO QCD prediction of ~\cite{NLOQCD}. The 
theoretical computation tends to be slightly lower than the measurements 
at high  $p_T$ values although staying compatible with them within errors.

\section{Discussion of results}

Our measurement of the $d\sigma/dp_T$ cross-section of the inclusive production
of hadrons in $\gamma^*\gamma^*$ interactions appears to agree well with both 
PYTHIA and NLO QCD predictions.

The L3 experiment has performed a similar analysis ~\cite{L3} and has observed 
that the $p_T$ spectrum of charged hadrons is slightly below the PYTHIA MC 
prediction while the derived $d\sigma/dp_T$ cross-section considerably exceeds 
the NLO QCD prediction at high $p_T $ values. 
We have repeated our analysis, adopting a ``L3-like'' set of selection criteria
which, compared to ours, corresponds to a less tight $W_{vis}$ cut 
($W_{vis}<$~78 GeV{\it/$c^2$} instead of 35 GeV{\it/$c^2$}) and 
a higher threshold 
of the total number of particles including neutrals (5 instead of 4). The 
looser $W_{vis}$ cut has the effect of increasing significantly the  
\mbox{e$^+$e$^- \rightarrow q \bar{q}~(\gamma)$} background (see Fig.~1)
which now dominates at large $p_T$ values. The resulting $dN/dp_T$ spectrum of 
charged particles for the ``L3-like'' events is presented in Fig.~5 together 
with the contributing channels. One observes an excess of data over the 
PYTHIA MC prediction. This disagreement between MC and data is likely to come 
from charged particles of background channels as these are introduced in much 
larger quantities than charged particles from 
\mbox{$\gamma^*\gamma^* \rightarrow \ hadrons$}, when the 
$W_{vis}$ cut is relaxed up to 78 GeV{\it/$c^2$}, as can be checked by comparing
Fig.~5 with Fig.~2. It legitimates, {\it $a ~posteriori$}, our  
$W_{vis}<$~35 GeV{\it/$c^2$} cut to minimize the contamination of charged 
particles from background channels.

The OPAL experiment has measured the differential $d\sigma/dp_T$ cross-section 
of the inclusive production of charged hadrons ~\cite{OPAL} for different 
intervals of $W$, the hadronic invariant mass corrected for detector effects. 
In the (10 GeV{\it/$c^2$} $< W <$~30 GeV{\it/$c^2$}) interval, the 
cross-section is compatible with the NLO prediction.  

\section{Conclusions}
The study of  the inclusive charged hadron production in two-photon collisions 
has been carried out at the DELPHI detector at LEP II. Measurements of the 
inclusive single-particle $p_T$ distribution and of the 
differential inclusive $d\sigma/dp_T$ cross-section have been extracted. 
The differential inclusive $d\sigma/dp_T$ cross-section is found to be 
compatible, within errors, with the PYTHIA and NLO QCD predictions up to 
high $p_T$, although systematic uncertainties limit the accuracy of the comparison
in this region. It is shown that if cuts such as those used in ~\cite{L3} are applied, 
$q \bar{q}$ background dominates at large $p_T$, making it difficult to 
draw conclusions on two-photon processes.

\subsection*{Acknowledgements}
\vskip 3 mm
We are greatly indebted to our technical 
collaborators, to the members of the CERN-SL Division for the excellent 
performance of the LEP collider, and to the funding agencies for their
support in building and operating the DELPHI detector.\\
We acknowledge in particular the support of \\
Austrian Federal Ministry of Education, Science and Culture,
GZ 616.364/2-III/2a/98, \\
FNRS--FWO, Flanders Institute to encourage scientific and technological 
research in the industry (IWT) and Belgian Federal Office for Scientific, 
Technical and Cultural affairs (OSTC), Belgium, \\
FINEP, CNPq, CAPES, FUJB and FAPERJ, Brazil, \\
Ministry of Education of the Czech Republic, project LC527, \\
Academy of Sciences of the Czech Republic, project AV0Z10100502, \\
Commission of the European Communities (DG XII), \\
Direction des Sciences de la Mati$\grave{\mbox{\rm e}}$re, CEA, France, \\
Bundesministerium f$\ddot{\mbox{\rm u}}$r Bildung, Wissenschaft, Forschung 
und Technologie, Germany,\\
General Secretariat for Research and Technology, Greece, \\
National Science Foundation (NWO) and Foundation for Research on Matter (FOM),
The Netherlands, \\
Norwegian Research Council,  \\
State Committee for Scientific Research, Poland, SPUB-M/CERN/PO3/DZ296/2000,
SPUB-M/CERN/PO3/DZ297/2000, 2P03B 104 19 and 2P03B 69 23(2002-2004),\\
FCT - Funda\c{c}\~ao para a Ci\^encia e Tecnologia, Portugal, \\
Vedecka grantova agentura MS SR, Slovakia, Nr. 95/5195/134, \\
Ministry of Science and Technology of the Republic of Slovenia, \\
CICYT, Spain, AEN99-0950 and AEN99-0761,  \\
The Swedish Research Council,      \\
The Science and Technology Facilities Council, UK, \\
Department of Energy, USA, DE-FG02-01ER41155, \\
EEC RTN contract HPRN-CT-00292-2002. \\


%

\newpage
\begin{table}[p]
\begin{center}
\begin{tabular}{|c|c|c|c|}
 \hline
   $p_T$, GeV{\it/c} & $<p_T>$, GeV{\it/c} &
\multicolumn{2}{|c|} {$d\sigma/dp_T$, pb/GeV{\it/c}}  \\
\hline 
 & & $|\eta|<1$ & $|\eta|<1.5$    \\
\hline 
   &       &           &      \\
1.6 - 2.0 & 1.76 & (2.36$\pm 0.02^{+0.88}_{-0.41}$) $\times 10^2$ \ \    & (3.00$\pm0.02^{+0.42}_{-0.60}$) $\times 10^2$ \ \  \\
\hline
    &      &           &      \\
2.0 - 2.4 & 2.17 & (8.98$\pm0.11^{+3.24}_{-1.18}$) $\times 10^1$ \ \    & (1.15$\pm0.01^{+0.09}_{-0.17}$) $\times 10^2$ \ \   \\
\hline
     &     &           &      \\
2.4 - 2.8 & 2.58 &(4.05$\pm0.07^{+1.30}_{-0.58}$) $\times 10^1$ \ \   & (5.23$\pm0.08^{+0.27}_{-0.82}$) $\times 10^1$ \ \   \\
\hline
     &     &           &      \\
2.8 - 3.2 & 2.98 &(2.10$\pm0.05^{+0.82}_{-0.27}$) $\times 10^1$ \ \   & (2.66$\pm0.06^{+0.30}_{-0.38}$) $\times 10^1$ \ \   \\
\hline
     &     &           &      \\
3.2 - 3.6 & 3.38 &(1.24$\pm0.04^{+0.44}_{-0.17}$) $\times 10^1$ \ \   & (1.61$\pm0.05^{+0.05}_{-0.25}$) $\times 10^1$  \ \  \\
\hline
      &    &           &      \\
3.6 - 4.0 & 3.78 &(7.31$\pm0.34^{+2.92}_{-1.06}$) \ \ \ \ \ \ \ \ \    & (9.41$\pm0.35^{+1.03}_{-1.69}$) \ \ \ \ \ \ \ \ \   \\
\hline
      &    &           &      \\
4.0 - 4.4 & 4.18 &(4.29$\pm0.26^{+2.07}_{-0.47}$) \ \ \ \ \ \ \ \ \    & (5.54$\pm0.27^{+0.85}_{-0.54}$)  \ \ \ \ \ \ \ \ \  \\
\hline
      &    &           &      \\
4.4 - 4.8 & 4.59 &(2.95$\pm0.22^{+1.36}_{-0.46}$) \ \ \ \ \ \ \ \ \    & (3.89$\pm0.24^{+0.42}_{-0.47}$)  \ \ \ \ \ \ \ \ \  \\
\hline
      &    &           &      \\
4.8 - 5.2 & 4.99 &(2.22$\pm0.19^{+1.05}_{-0.12}$) \ \ \ \ \ \ \ \ \    & (2.78$\pm0.20^{+0.29}_{-0.10}$)  \ \ \ \ \ \ \ \ \  \\
\hline
      &    &           &      \\
5.2 - 5.6 & 5.39 &(1.33$\pm0.16^{+0.62}_{-0.05}$)  \ \ \ \ \ \ \ \ \   & (1.65$\pm0.16^{+0.19}_{-0.06}$)  \ \ \ \ \ \ \ \ \  \\
\hline
      &    &           &      \\
5.6 - 6.0 & 5.79 &(1.36$\pm0.17^{+0.41}_{-0.25}$)  \ \ \ \ \ \ \ \ \   & (1.70$\pm0.19^{+0.12}_{-0.24}$)  \ \ \ \ \ \ \ \ \  \\
\hline
      &    &           &      \\
6.0 - 6.4 & 6.19 &(9.70$\pm1.42^{+4.04}_{-1.20}$) $\times 10^{-1}$ & (1.16$\pm0.15^{+0.15}_{-0.14}$)  \ \ \ \ \ \ \ \ \   \\
\hline
      &    &           &      \\
6.4 - 6.8 & 6.59 &(4.57$\pm1.01^{+3.26}_{-0.88}$) $\times 10^{-1}$ & (8.34$\pm1.36^{+0.47}_{-2.66}$) $\times 10^{-1}$ \\
\hline
      &    &           &      \\
6.8 - 7.2 & 6.98 &(5.44$\pm1.11^{+5.96}_{-3.03}$) $\times 10^{-1}$ & (6.65$\pm1.12^{+2.52}_{-2.90}$) $\times 10^{-1}$ \\
\hline
      &    &           &      \\
7.2 - 7.6 & 7.38 &(5.13$\pm1.04^{+1.18}_{-0.92}$) $\times 10^{-1}$ & (5.43$\pm1.09^{+0.28}_{-0.23}$) $\times 10^{-1}$ \\
\hline
      &    &           &      \\
7.6 - 8.0 & 7.78 &(2.93$\pm0.91^{+1.70}_{-1.57}$) $\times 10^{-1}$ & (3.67$\pm0.92^{+0.38}_{-1.42}$) $\times 10^{-1}$ \\
\hline
      &    &           &      \\
8.0 - 9.0 & 8.44 &(1.56$\pm0.68^{+3.48}_{-1.33}$) $\times 10^{-1}$ & (2.65$\pm1.23^{+1.94}_{-2.30}$) $\times 10^{-1}$ \\
\hline
      &    &           &      \\
9.0 - 10.0 & 9.47 &(1.08$\pm0.59^{+1.76}_{-0.89}$) $\times 10^{-1}$ & (1.71$\pm0.86^{+1.41}_{-1.30}$) $\times 10^{-1}$ \\
\hline
      &    &           &      \\
10.0 - 12.0 & 10.87 &(0.53$\pm0.22^{+1.68}_{-0.44}$) $\times 10^{-1}$ & (0.68$\pm0.28^{+1.37}_{-0.49}$) $\times 10^{-1}$ \\
\hline
      &    &           &      \\
12.0 - 16.0 & 13.53 &(0.16$\pm0.05^{+0.26}_{-0.02}$) $\times 10^{-1}$ & (0.23$\pm0.07^{+0.43}_{-0.14}$) $\times 10^{-1}$ \\
\hline
\end{tabular}
\vspace*{1.cm}
\caption{Differential inclusive $d\sigma/dp_T$ of charged particles produced
in \mbox{$\gamma^*\gamma^* \rightarrow \ hadrons$} collisions,
for $|\eta|<$1, $|\eta|<$1.5 and $p_T>$~1.6 GeV{\it/c}. The first error 
is statistical, the second is the systematic uncertainty. The data are 
background subtracted and corrected for detector inefficiency and selection
cuts.}
\end{center}
\end{table}
\vfill 

\newpage
\begin{figure}[p]
\epsfxsize=\textwidth
\begin{center}
\mbox{\epsfig{figure=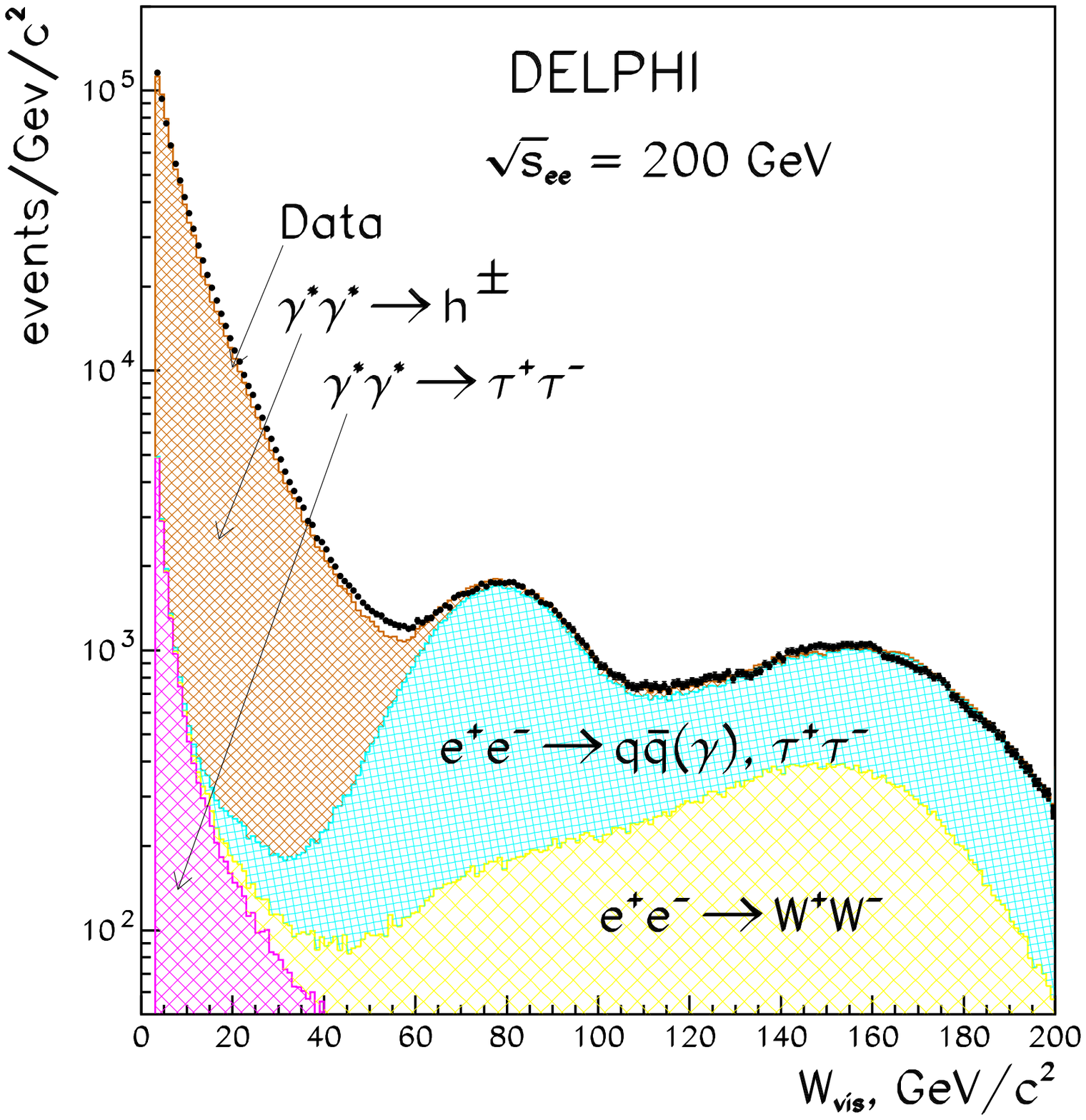,width=16. cm}}
\end{center}
\caption{$W_{vis}$  distributions for the data and for the simulated 
\mbox{$\gamma^*\gamma^* \rightarrow \ hadrons$} (medium cross-hatching),
\mbox{$\gamma^*\gamma^* \rightarrow \tau^+ \tau^-$} (second largest cross-hatching) ,
${\mathrm e}^+{\mathrm e}^-\rightarrow q \bar{q}~(\gamma), \tau^+ \tau^-$ 
(small cross-hatching) and
${\mathrm e}^+{\mathrm e}^-\rightarrow {\mathrm W}^+ {\mathrm W}^-$ 
(largest cross-hatching) events at 
$\sqrt{s_{{\mathrm e}{\mathrm e}}}=200$ GeV.}
\end{figure}
%
\begin{figure*}[p]
\epsfxsize=\textwidth
\begin{center}
\mbox{\epsfig{figure=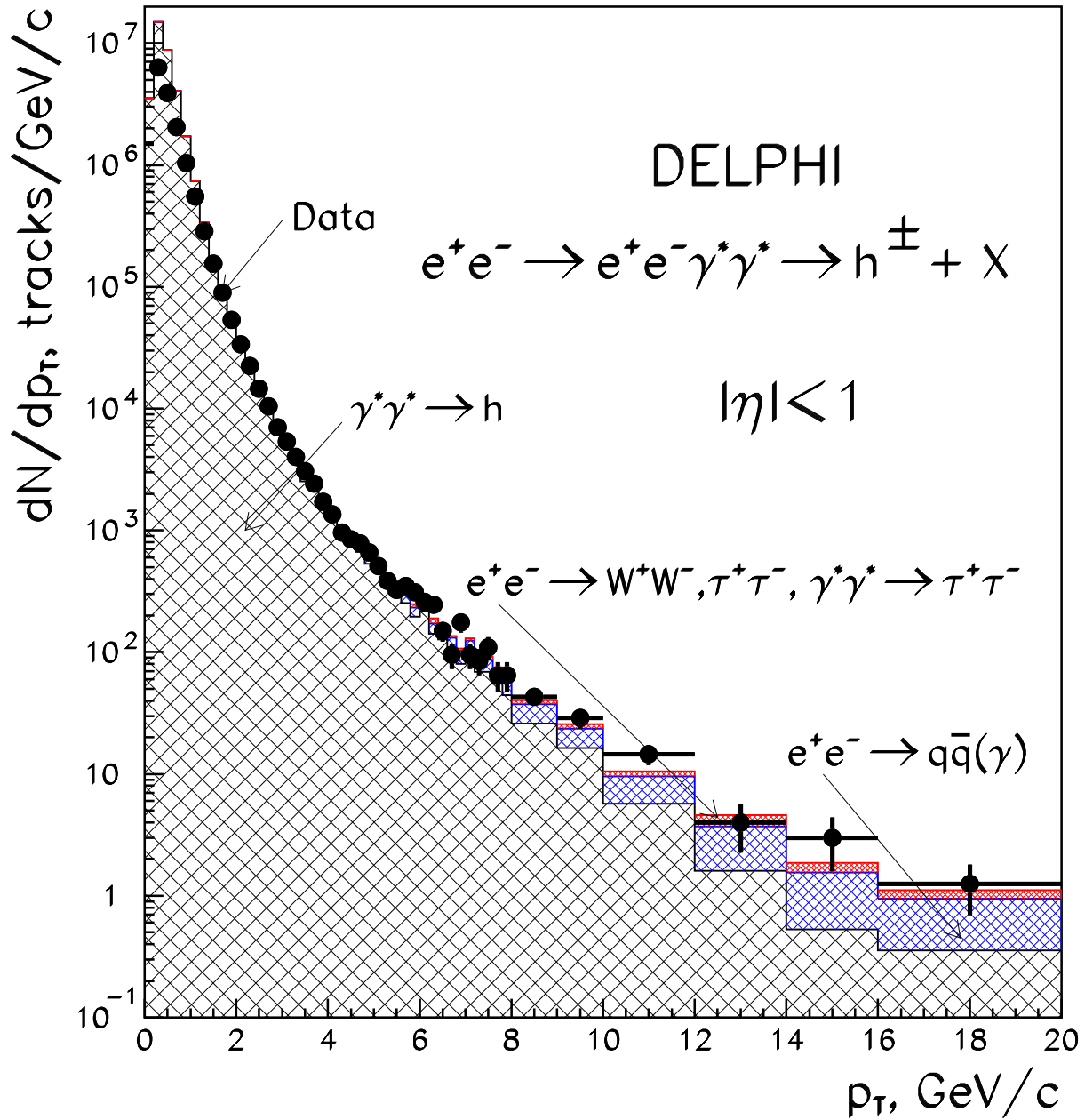,width=16. cm}}
\end{center}
\caption{$p_T$ distribution of charged particles of the selected sample of 
events, for $|\eta|<$1 together with the Monte Carlo generated contributing 
processes:
\mbox{$\gamma^*\gamma^* \rightarrow \ hadrons$} (largest cross-hatching),
${\mathrm e}^+{\mathrm e}^-\rightarrow q \bar{q}~(\gamma)$ (medium cross-hatching), 
${\mathrm e}^+{\mathrm e}^-\rightarrow {\mathrm W}^+ {\mathrm W}^-$,
$\tau^+ \tau^-$,   
\mbox{$\gamma^*\gamma^* \rightarrow$ $\tau^+ \tau^-$} 
(small cross-hatching).}
\end{figure*}
%
\begin{figure*}[p]
\epsfxsize=\textwidth
\begin{center}
\mbox{\epsfig{figure=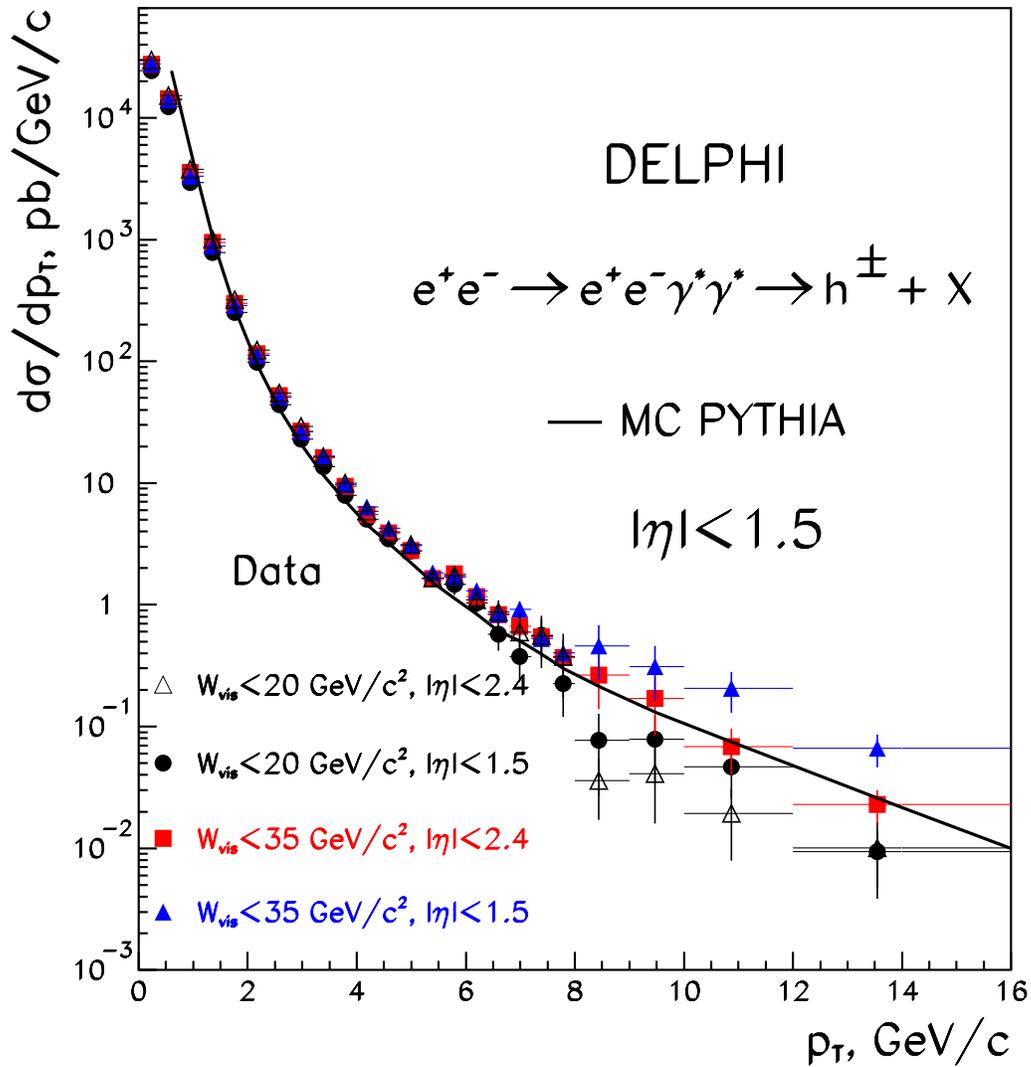,width=16. cm}}
\end{center}
\caption{Differential inclusive $d\sigma/dp_T$ distributions of charged 
particles with $|\eta|<$1.5, produced in $\gamma^*\gamma^*$ collisions, for 
different sets of initial selection criteria. (The lower limit of $W_{vis}$ 
was $W_{vis}>$~5 GeV{\it/$c^2$}).
The data are background subtracted and corrected for detector inefficiency 
and selection cuts. The line is the corresponding 
PYTHIA prediction for \mbox{$\gamma^*\gamma^* \rightarrow \ hadrons$}.}
\end{figure*}
%
\begin{figure*}[p]
\epsfxsize=\textwidth
\begin{center}
\mbox{\epsfig{figure=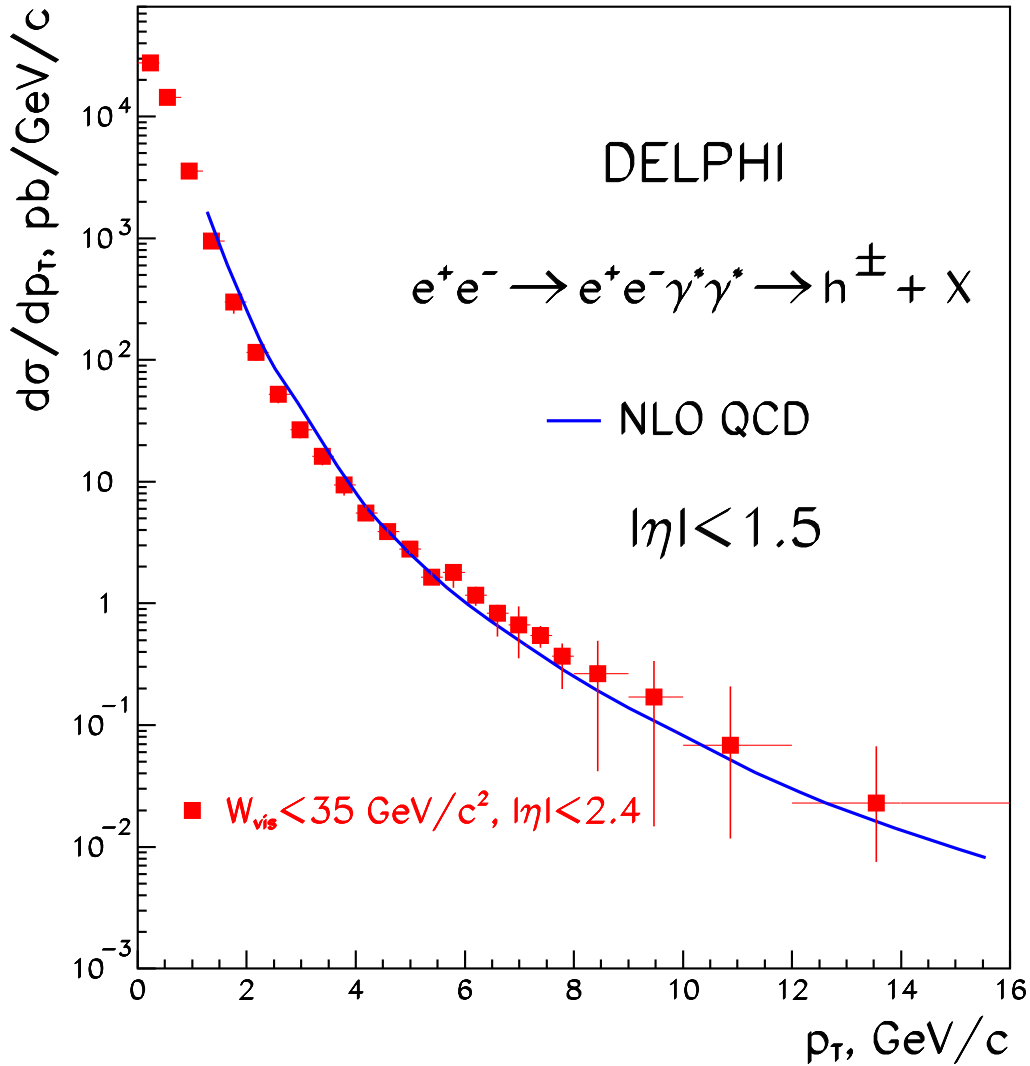,width=16. cm}}
\end{center}
\caption{Differential inclusive $d\sigma/dp_T$ distribution of charged 
particles produced in $\gamma^*\gamma^*$ collisions for $|\eta|<$1.5
and \mbox{5 GeV{\it/$c^2$}$< W_{vis} < $ 35 GeV{\it/$c^2$}}. 
The original data sample used to extract this cross section included
 tracks with $10^{\circ} \ < \ \theta \ < \ 170^{\circ} ~~(|\eta|<2.4)$.
The data are shown as points with 
statistical + systematical error bars. They are background subtracted and 
corrected for detector inefficiency and selection cuts. The line is the 
NLO QCD prediction of~\cite{NLOQCD} for 
\mbox{$\gamma^*\gamma^* \rightarrow \ hadrons$}.}
\end{figure*}
%
\begin{figure*}[p]
\epsfxsize=\textwidth
\begin{center}
\mbox{\epsfig{figure=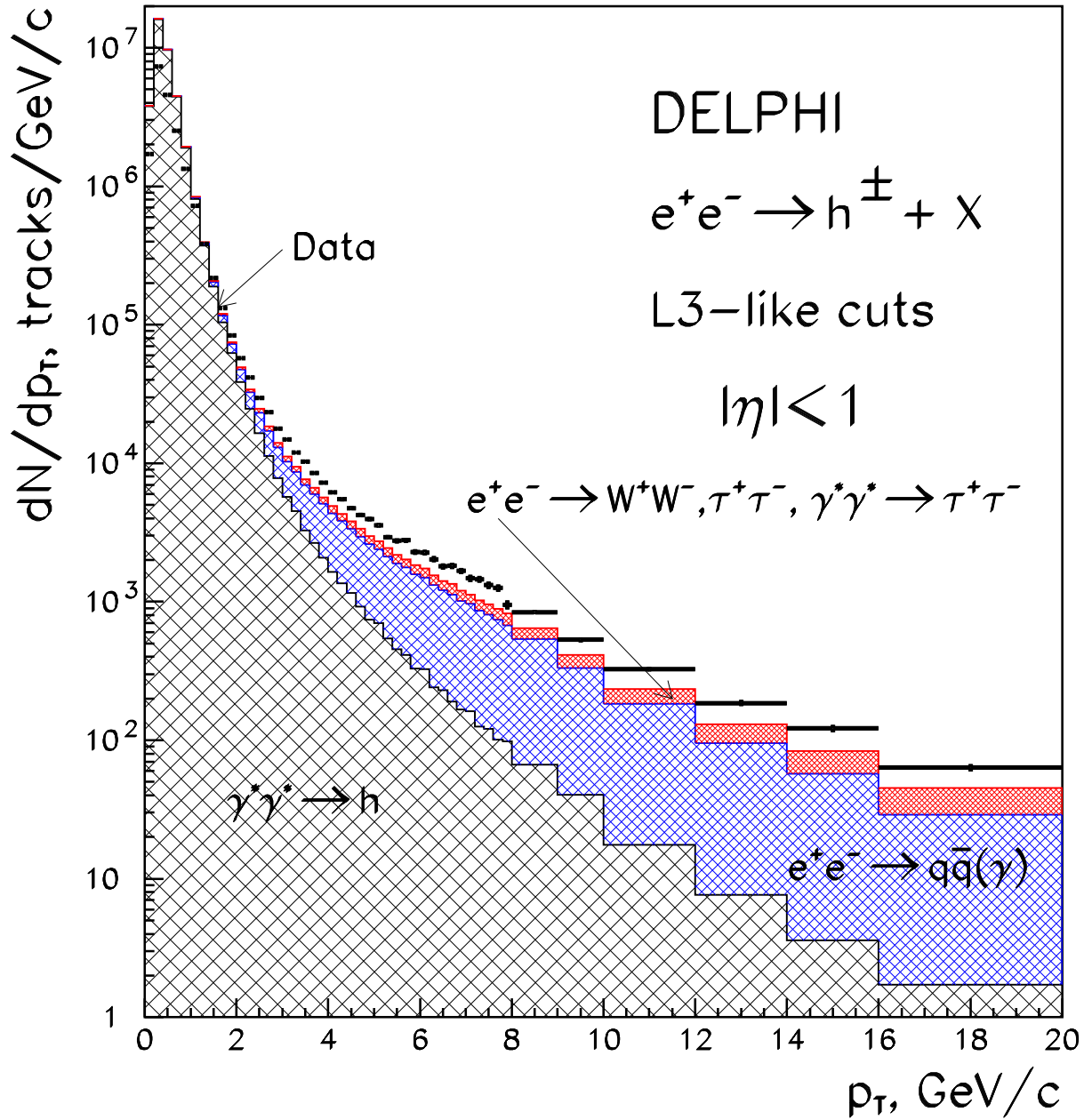,width=16. cm}}
\end{center}
\caption{$p_T$ distribution of charged particles of the event sample after
application of the ``L3-like'' selection criteria, for $|\eta|<$1
and \mbox{5 GeV{\it/$c^2$}$< W_{vis} < $ 78 GeV{\it/$c^2$}}, together 
with the Monte Carlo generated contributing processes:
\mbox{$\gamma^*\gamma^* \rightarrow \ hadrons$} (largest cross-hatching),
${\mathrm e}^+{\mathrm e}^-\rightarrow q \bar{q}~(\gamma)$ (medium cross-hatching), 
${\mathrm e}^+{\mathrm e}^-\rightarrow {\mathrm W}^+ {\mathrm W}^-$,
$\tau^+ \tau^-$,   
\mbox{$\gamma^*\gamma^* \rightarrow$ $\tau^+ \tau^-$} 
(small cross-hatching).}
\end{figure*}

\end{document}